# Neurovascular coupling: insights from multi-modal dynamic causal modelling of fMRI and MEG

Amirhossein Jafarian *[1], Vladimir Litvak[1], Hayriye Cagnan[2,3], Karl J. Friston[1], Peter Zeidman[1]

[1] The Wellcome Centre for Human Neuroimaging, University College London, United Kingdom

[2] MRC Brain Network Dynamics Unit (BNDU) at the University of Oxford, Oxford, UK

[3] Nuffield Department of Clinical Neurosciences, University of Oxford, Oxford, UK

**Correspondence:** Amirhossein Jafarian

The Wellcome Centre for Human Neuroimaging,

12 Queen Square, London, UK. WC1N 3AR

*E-mail*: a.jafarian@ucl.ac.uk

## Abstract

This technical note presents a framework for investigating the underlying mechanisms of neurovascular coupling in the human brain using multi-modal magnetoencephalography (MEG) and functional magnetic resonance (fMRI) neuroimaging data. This amounts to estimating the evidence for several biologically informed models of neurovascular coupling using variational Bayesian methods and selecting the most plausible explanation using Bayesian model comparison. First, fMRI data is used to localise active neuronal sources. The coordinates of neuronal sources are then used as priors in the specification of a DCM for MEG, in order to estimate the underlying generators of the electrophysiological responses. The ensuing estimates of neuronal parameters are used to generate neuronal drive functions, which model the pre- or post-synaptic responses to each experimental condition in the fMRI paradigm. These functions form the input to a model of neurovascular coupling, the parameters of which are estimated from the fMRI data. This establishes a Bayesian fusion technique that characterises the BOLD response – asking, for example, whether instantaneous or delayed pre- or post-synaptic signals mediate haemodynamic responses. Bayesian model comparison is used to identify the most plausible hypotheses about the causes of the multimodal data. We illustrate this procedure by comparing a set of models of a single-subject auditory fMRI and MEG dataset. Our exemplar analysis suggests that the origin of the BOLD signal is mediated instantaneously by intrinsic neuronal dynamics and that neurovascular coupling mechanisms are region-specific. The code and example dataset associated with this technical note are available through the statistical parametric mapping (SPM) software package.

# 1. Introduction

To interpret the blood oxygenation-level dependent (BOLD) contrast, and its disruption due to aging, disease or pharmacological interventions, a better understanding of the mechanisms of neurovascular coupling is needed. Neuronal activity triggers vasodilation, both directly via signalling molecules such as nitric oxide (Li and Iadecola, 1994) and indirectly via astrocytes (Takano et al., 2006). The ensuing change in blood flow is accompanied by a change in blood oxygenation (Logothetis, 2001; Filosa et al 2007), detectable as the BOLD contrast. However, there are many questions yet to be answered about the origin of the BOLD response in the human brain (Arthurs et al, 2002; Hall et al., 2016). For instance, is it driven by pre- or post-synaptic potentials of neuronal populations? Does a region's BOLD response depend on local or distal neuronal projections? What causes region-specific differences in the BOLD response?

Invasive recordings in animal models are commonly employed to distinguish neuronal, vascular and haemodynamic contributions to the BOLD response (e.g. Logothetis et al., 2001; Grill-Spector et al., 2006; Snyder et al., 2016). However, the same imaging techniques cannot be adopted to study the human brain *in vivo*, which necessitates the use of non-invasive functional imaging. BOLD contrast imaging using fMRI provides high spatial resolution for localising activity and, with suitable models, enables inferences to be made about the mechanisms of neurovascular coupling (Stephan et al. 2007). This imaging technique typically has greater temporal resolution than other MRI methods used to study neurovascular coupling, such as arterial spin labelling; however, it is still too slow to inform detailed models of neuronal activity. By contrast, electromagnetic recordings such as MEG provide exquisite temporal resolution – at the level of electrophysiological dynamics – which in turn support the identification of detailed neural models (David et al., 2006). The question then arises: how can we leverage the sensitivity of fMRI to haemodynamics and MEG to neuronal dynamics to best study neuro-vascular interactions in humans non-invasively?

The approach pursued here, building on several previous studies, is to combine a detailed neuronal model fitted to EEG or MEG data with a model of neurovascular coupling and haemodynamics fitted to fMRI data. Our objective was to develop a procedure, with associated tools implemented in the SPM software package, for investigating neurovascular coupling using multi-modal data. To illustrate the methods and ground them with an empirical example, we analysed a dataset in which a single subject performed an auditory (roving oddball) task, while undergoing MEG and fMRI on separate days.

To model the example multi-modal data, our first consideration was which neuronal model to use. Neuronal models of varying complexity have been used in previous studies examining neurovascular coupling. For example, Riera et al. (2005, 2006, 2007) explored mechanisms of neurovascular coupling using fMRI- EEG data. In their models, the BOLD response could be induced by pre- and/or post-synaptic potentials associated with a single population of deep pyramidal cells, connected with two populations of inhibitory interneurons. Voges et al. (2012) investigated neurovascular coupling in the context of epilepsy, using a neural mass model with one inhibitory and one excitatory sub-population, based on Wendling et al. (2000, 2005) and Jansen and Rit (1995). A recent study by Friston et al. (2017) used a four population canonical microcircuit (CMC) model (Bastos et al., 2012) to demonstrate that fMRI and EEG/LFP data features may be uncorrelated, despite having the same underlying neuronal sources. They coupled the CMC model, which includes superficial and deep pyramidal cells as well as excitatory and inhibitory neurons, with the extended Balloon model typically employed in DCM for fMRI (Stephan et al. 2007). This combined model, so far demonstrated only with simulated EEG / LFP data, has the potential to reveal laminar specific contributions to the BOLD response. For this reason, we used the CMC model here, although it could easily be replaced with any other neural mass model.

Our second consideration was the form of the model of neurovascular coupling model and which neuronal sources should drive it. Previous studies have explored detailed neurovascular coupling models using non-invasive measurements (see review by Huneau et al., 2015). For example, Sotero and Trujllo-Baretto (2007) proposed a model in which lumped excitatory and inhibitory neuronal inputs drive a detailed model of metabolic change and haemodynamics. Other models have been evaluated by Rosa et al. (2012), who embedded the forward model proposed by Riera et al. (2006) in a (variational) Bayesian framework. They performed a Bayesian model comparison to evaluate different neuro-vascular coupling functions based on synaptic activity and / or post-synaptic firing rates. Here, we took a similar approach and compared the evidence for different combinations of pre- or post-synaptic neuronal inputs, as well as exogenous inputs from different neuronal populations, using Bayesian model comparison. These mixtures of neuronal activity entered an established neurovascular coupling model (Friston et al., 2000) in which a vasodilatory signal, thought to be nitric oxide, induces flow and is subject to feedback induced by that flow. This in turn drives haemodynamics, captured by the balloon model (Buxton et al., 1998), and in turn a model of the fMRI signal (Stephan et al. 2007). We emphasise, however, that in the analysis procedure we set out, any of these components could be substituted or compared based on their model evidence.

Our third consideration was how to integrate MEG and fMRI data to efficiently estimate the parameters of the neuronal, neurovascular and haemodynamic parts of the model. To make inversion tractable, reasonable independence assumptions can be made about the parameters (i.e. a mean-field approximation). For example, Rosa et al. (2012) used a three-step variational Bayesian estimation procedure, where they first estimated neuronal parameters, then neurovascular coupling parameters, and finally the parameters governing haemodynamics. Here, we also used variational Bayesian inference methods, and divided the estimation into a neuronal part and a neurovascular / haemodynamics part, linked by *neuronal drive functions*. These functions are canonical synaptic responses to each experimental condition from each neuronal population, derived from a neural mass model which has been fitted to the MEG data. These functions then form the input to the neurovascular coupling model, which in turn drives the haemodynamics. Parameters relating to the neurovascular and haemodynamic parts of the model are estimated from the fMRI data. This approach offers convenience and flexibility, because the neuronal drive functions can be generated from any of the neural mass models available in the DCM framework without the need for re-implementation.

In summary, the framework we set out in this paper couples a dynamic causal model of laminar specific neuronal responses (Bastos et *al.*, 2012) with a model of neurovascular coupling and the BOLD response (Stephan, et *al.*, 2007). They are linked by neuronal drive functions, which model the pre- or post-synaptic activity of each neuronal population under each experimental condition. The form of the neuronal drive or coupling functions is parameterised to enable hypothesis testing using Bayesian model comparison. To illustrate the approach, we specified a factorial model space covering a number of foundational questions about the mechanisms of neurovascular coupling. The factors were: presynaptic versus postsynaptic contributions to the neurovascular signal, whether the inputs to neurovascular coupling were region-specific, whether distal regions contributed to local changes in BOLD contrasts, and whether neurovascular delays associated with the release of vasoactive agents (e.g. calcium) should be modelled. This model space allowed us to perform a series of family-wise model comparisons, quantifying the evidence for each question in turn.

This paper has five sections. In section two, we set out the theory underlying the approach. In section three, we introduce the example dataset and detail the specifics of the model specification for these data. In section four, the results of the example analyses are illustrated. Finally, the conclusions and future applications are considered in section five.

## 2. Theory

### 2.1 Dynamic Causal Modelling for MEG

A biologically informed generative model of multimodal fMRI and MEG data is shown in Figure 1. This DCM illustrates the common underlying neuronal generators of both MEG and fMRI measurements, mediated by a spatial lead field and BOLD response model, respectively. We will explain each part of the model in the following sections, before illustrating its application to real data. All variables are defined in tables 1-4.

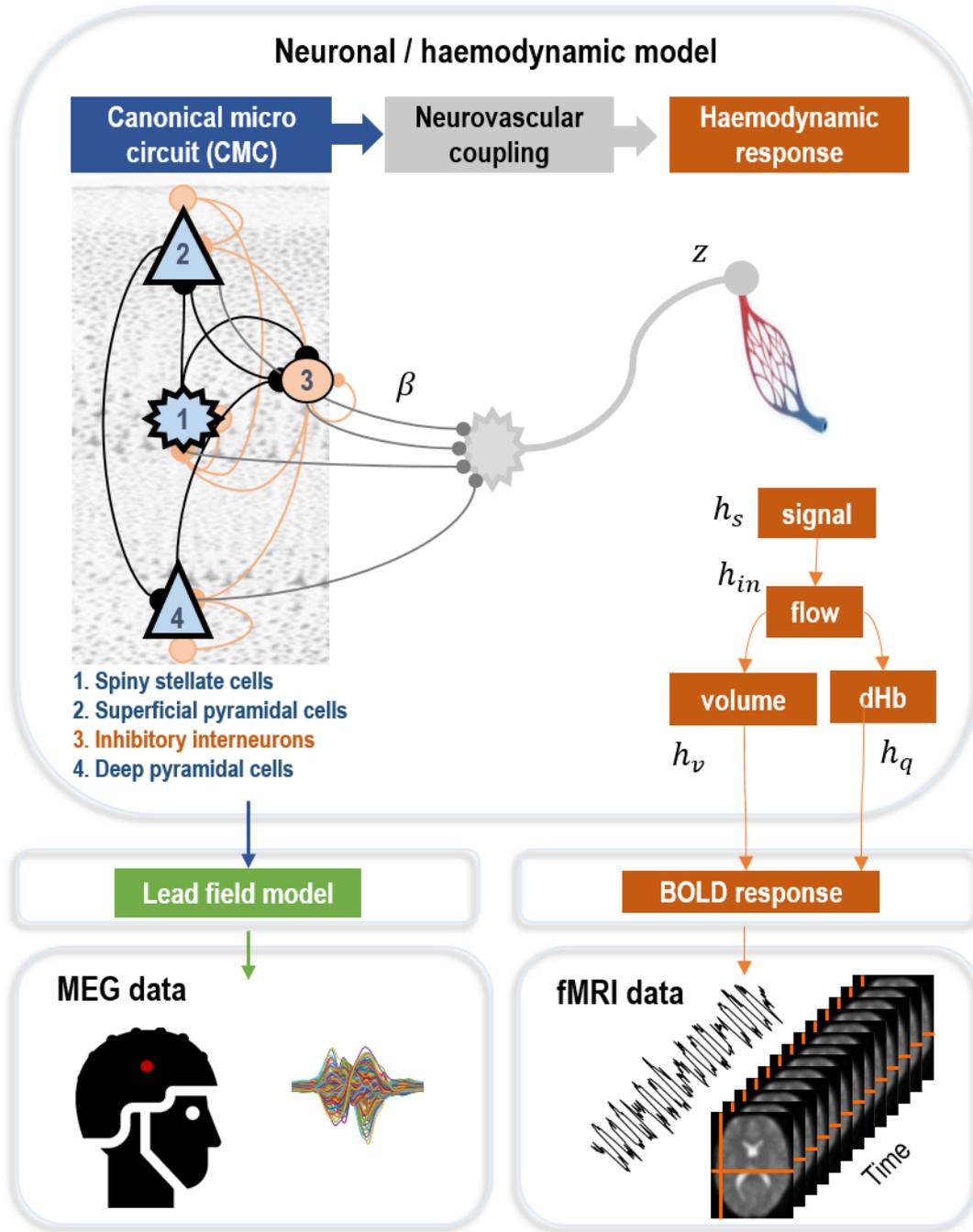

Figure 1. Components of a forward model of fMRI and electrophysiological data. Top: laminar specific canonical microcircuit (CMC) comprising four populations (numbered 1-4) per brain region. Each CMC is linked through extrinsic (between region) forward and backward connections. Pre- or postsynaptic neuronal signals are combined (at the level of the putative astrocytes), and this drives the haemodynamic part of the model. Blood flow is increased to the venous compartment (pictured), which is accompanied by changes of blood volume and the level of deoxyhemoglobin. Bottom: Electrophysiological and fMRI measurements arise from the neuronal and haemodynamic parts of the model respectively, mediated by a spatial lead field model for MEG and a BOLD signal model for fMRI. To make inversion of this model tractable, we split the neuronal and haemodynamic parts and connected them via neuronal response functions – see text and Figure 5.

### 2.1.1 Generative model of neuronal responses

We used the canonical microcircuit (CMC), which models the circuitry of a typical cortical column (Bastos *et al.*, 2012; Douglas and Martin, 1991). It comprises four neuronal populations per brain region: spiny stellate cells in the granular layer (ss), superficial pyramidal cells in the supragranular layer (sp), inhibitory interneurons distributed in all layers of the cortex (ii) and deep pyramidal cells in the infragranular layers (dp), as shown in Figure 1. The four populations within each cortical column have intrinsic (inter-and intra-laminar) connections that are ubiquitous in most cortical areas (Thomson and Bannister, 2003; Binzegger et al., 2004; Haeusler and Maass, 2007). Experimental and extrinsic inputs are received by spiny stellate cells in the granular layer (hereinafter referred to as extrinsic forward connections) that project to superficial pyramidal cells and thereafter to deep pyramidal cells. Each excitatory connection establishes reciprocal connections with inhibitory interneurons. All populations have a recurrent (self) inhibitory connection proportional to the level of excitation of the neuronal population. There are two types of external (extrinsic) input entering each microcircuit from different levels of the cortical hierarchy. Inputs can be bottom-up (forward) connections arising from superficial pyramidal cells of the level below, targeting spiny stellate cells and deep pyramidal cells. Alternatively, inputs can be top-down (backward) connections arising from deep pyramidal cells of the level above, targeting inhibitory interneurons and superficial pyramidal cells (Felleman and Van Essen, 1991; Hilgetag et al., 2000).

Two conversion operators govern the dynamics of each neuronal population (Jansen and Rit 1995). The first operator converts the mean pre-synaptic firing rate $m$ to the mean postsynaptic membrane potential $V$ as follows (Freeman, 1975):

$$V = h \otimes m \tag{1}$$

Where $\otimes$ denotes the linear convolution operator and $h$ is the impulse response function (synaptic kernel) with synaptic rate constant $\kappa$:

$$h(t) = \begin{cases} \dfrac{t}{\kappa} e^{-\frac{t}{\kappa}}, & x \geq 0 \\ 0, & x < 0 \end{cases} \tag{2}$$

The second operator then transforms the postsynaptic membrane potential into a firing rate, which forms the input to the next connected neural population (Wilson & Cowan, 1972):

$$\sigma(V) = \frac{1}{1 + \exp(-V)} - \frac{1}{2} \tag{3}$$

The dynamics of postsynaptic potentials in region $k$, population $i$, $V_i^k$, obey second order differential equations as follows:

$$\left(1 + \frac{1}{\kappa_i}\frac{d}{dt}\right)^2 V_i^K(t) = f_i(V_{ex}^\sigma, V_i^K, u) \tag{4}$$

where the intrinsic presynaptic excitations are given by $V_j^K$, the term $V_{ex}^\sigma$ denotes extrinsic drives of a population $\sigma$ in a distal region $ex$; and the function $f$ is defined as follows (Friston et al., 2017):

$$f_i(V_{ex}^\sigma, V_i^K, u)$$
$$= \begin{cases} A_f^{sp \to ss}\sigma(V_{sp}^{ex}) - a_{ss \to ss}\sigma(V_{ss}^k) - a_{sp \to ss}\sigma(V_{sp}^k) - a_{ii \to ss}\sigma(V_{ii}^k) + Cu_k & if\ i = ss \\ A_b^{dp \to sp}\sigma(V_{dp}^{ex}) - a_{sp \to sp}\sigma(V_{sp}^k) + a_{ss \to sp}\sigma(V_{ss}^k) - a_{ii \to sp}\sigma(V_{ii}^k) & if\ i = sp \\ A_b^{dp \to ii}\sigma(V_{dp}^{ex}) - a_{ii \to ii}\sigma(V_{ii}^k) - a_{dp \to ii}\sigma(V_{dp}^k) + a_{ss \to ii}\sigma(V_{ss}^k) + a_{sp \to ii}\sigma(V_{sp}^k) & if\ i = ii \\ A_f^{sp \to dp}\sigma(V_{sp}^{ex}) - a_{dp \to dp}\sigma(V_{dp}^k) - a_{ii \to dp}\sigma(V_{ii}^k) + a_{sp \to dp}\sigma(V_{sp}^k) & if\ i = dp \end{cases} \tag{5}$$

The laminar specificity of the extrinsic and intrinsic connections in equation (3) are specified by placing prior constraints on the intrinsic (within-region) connectivity parameters $a_{* \to *}$ as well as on the elements of the extrinsic (between-region) forward and backward adjacency matrices $A_{f,b}^{* \to *}$ ($A_f^{sp \to ss}$ and $A_f^{sp \to dp}$ denotes forward connections matrices, whereas backward connection matrices are specified by $A_b^{dp \to sp}$ and $A_b^{dp \to ii}$ matrices). Matrix $C$ parameterises the experimental driving input entering the system. These modelled neuronal dynamics are the common source of both the fMRI and MEG signals. As we will explain later, in DCM for MEG, we estimate condition specific forward and backward matrices $B_{f,b}$, which are applied (algebraically added) to the $A_{f,b}$ matrices and $a_{* \to *}$ parameters in order to model the differences between experimental conditions.

### 2.1.2 MEG observation model

The observation function for MEG data has the following form (Daunizeau et al., 2009):

$$y_{MEG} = \sum_k \Upsilon^k \Delta_0^k \sum_j \Psi_j v^j(t) + \epsilon_M \tag{6}$$

where $\epsilon_M \sim N(0, \sigma_M I)$ are I.I.D. measurement errors, $\Upsilon^k$ is a gain matrix for brain region $k$ and $\Delta_0^k$ is a Laplacian operator that is modelled as a mixture of spatial basis functions of the gain matrix as follows:

$$\Delta_0^k = \sum_n \Lambda_n^k \Theta_n^k \tag{7}$$

where $\Lambda_n^k$ are the spatial eigenvalues of the gain matrix and $\Theta_n^i$ are parameters to be estimated. The term $\sum_j \Psi_j v^j(t)$, where $j$ is the index of neuronal population, quantifies the contribution (modelled by unknown vector $\Psi_j$) of neuronal populations (denoted by $v^j(t)$) to the MEG signal). This completes the forward model of MEG data. Next, we detail the haemodynamic model used for fMRI data, before describing our novel approach to linking the two modalities.

## 2.2 Haemodynamic model
### 2.2.1 Generative model of neurovascular coupling

Neuronal dynamics (presynaptic or postsynaptic) excite neurovascular coupling mechanisms, which in turn trigger the vascular system to provide oxygen for neuronal consumption. While detailed models of the neurovascular system have been developed (e.g. Carmignoto and Gomez-Gonzalo 2010; Figley et al 2011), the lack of temporal resolution of fMRI places a limit on the complexity of models that can be inverted efficiently (Huneau et al., 2015; Pang et al., 2017). The framework set out in this paper provides the necessary tools for comparing the evidence for different models of neurovascular coupling. Two groups of models will be compared in this paper to illustrate the approach.

The first group of models posit that an instantaneous neurovascular response to neuronal activity (presynaptic firing rates or postsynaptic potentials) gives rise to the BOLD response. This is mediated by the release of nitric oxide (NO) and glutamate, which regulate and induce blood flow. The neurovascular signal can therefore be characterised as the algebraically scaled and summed responses associated with different neuronal populations. The scaling can either be considered to be the same for all regions, or different across regions, and we will compare the evidence for each of these options below. Additionally, we evaluated models where presynaptic inputs to each of the neuronal populations in the CMC were grouped into inhibitory, excitatory and extrinsic signals, each scaled by global coefficients (equal across regions) and summed to generate inputs to the haemodynamic model, as proposed in Friston et al (2017). Grouping the neuronal contributions in this way could offer a more parsimonious model than parameterising every neuronal population's contribution. In this study, all scale values associated with the neurovascular parameters had a (relatively) flat prior, placing minimal constraints on their value.

Alternatively, there might be a delay between the neuronal activity and haemodynamic response, due to the kinetics of intercellular calcium levels in the collaterals of astrocytes (Bazargani and Attwell 2016). Therefore, a second class of neurovascular models was included with additional delay factors. A parsimonious model that captures the mean delay with time constant $\tau_n$ due to elevation of intercellular calcium level is governed by a second order linear system with an impulse response function proposed by Pang et al. (2017):

$$f_{nc}(t) = \begin{cases} \dfrac{t}{\tau_{nc}} e^{-\frac{t}{\tau_{nc}}}, & x \geq 0 \\ 0, & x < 0 \end{cases} \quad (8)$$

The prior expected value of the delay factor in equation 8 was $0.7\ s$, based on recent observations from animal studies (Masamoto et al., 2015).

### 2.2.2 Generative model of the BOLD response

The output of any of the neurovascular coupling models considered above is a vasodilatory signal that alters the blood flow and accordingly the blood volume and oxygenation level. The haemodynamic model explains the dynamics of the vascular system as follows (Friston et al.,2000 & 2003):

$$\begin{aligned}
\dot{h}_s &= z - \eta h_s - \chi(h_{in} - 1) \\
\dot{h}_{in} &= h_s \\
\dot{h}_v &= \frac{1}{\tau_h}(h_{in} - h_v^{\frac{1}{\alpha}}) \\
\dot{h}_q &= \frac{1}{\tau_h}(h_{in} \frac{1 - (1 - E_0)^{\frac{1}{h_{in}}}}{E_0} - h_v^{\frac{1}{\alpha}} \frac{h_q}{h_v})
\end{aligned} \quad (9)$$

The first two lines in equation (3) are a damped filter (with the resonance frequency of the vasomotor signal, i.e. 0.1 Hz) that convert the neurovascular signal, $z$, to a vasodilatory signal $h_s$. The parameters $\eta$ and $\chi$ in the first equation are the decay rates of the vasodilatory signal and the auto-regulatory feedback term, respectively. Activation of the vasodilatory signal causes alteration in blood inflow $h_{in}$ to the venous compartments, which in turn causes an increase in blood volume $h_v$ and a reduction in the level of deoxyhaemoglobin $h_q$. The model for blood perfusion dynamics is given by Buxton et al.'s (2004,1998) Balloon model in the third and fourth lines in equation (9). The mean rate constant $\tau_h$ in the Balloon model is the time taken for blood to pass through the venous compartment (the transit time). The parameter for the blood vessel stiffness is $\alpha$ and is known as Grubb's coefficient, and $E_0$ is the net oxygen extraction fraction at rest, which characterises the fMRI baseline.

### 2.2.3 fMRI observation model

Finally, the change in blood volume and deoxyhaemoglobin combine to generate the BOLD signal:

$$y_{BOLD} = V_0 \left\{ k_1 \cdot (1 - h_q) + k_2 \cdot \left(1 - \frac{h_q}{h_v}\right) + k_3 \cdot (1 - h_v) \right\} + \epsilon_B \qquad (10)$$

With the addition of noise, this is the BOLD signal measured in the scanner. It comprises of physiological and field sensitive parameters, listed in Table 3.

## 2.3 Multimodal estimation procedure

The parts of the model described so far specify a pathway from neuronal activity to MEG and fMRI signals. In this section we set out a novel method for combining these model components and estimating their parameters. The procedure has three stages. First, a typical mass-univariate SPM analysis is performed on the fMRI data, to locate brain regions that evince experimental effects. Second, a DCM for MEG is specified, comprising a neuronal part (Section 2.1.1) and an observation part (Section 2.1.2). The coordinates of the brain regions identified in the fMRI analysis are used as prior constraints on the observation part, which projects neuronal activity to the scalp surface. A DCM is then fitted to the MEG data using the standard variational Laplace scheme (Friston, 2007), which provides an estimate of the parameters and the log model evidence (approximated by the negative variational free energy). Next, using the posterior expectations of the neuronal parameters, the DCM is used to generate a posterior predictive neuronal response to each experimental condition; hereafter, *neuronal drive functions*, which form a bridge between the MEG model and the fMRI model.

To clarify this approach, let the simulated electrophysiological response (e.g., pre or post synaptic signals) of population $i$ in region $j$ for the $1, \ldots, nu$ conditions be denoted by $f_1^{ij}(t), \ldots, f_{nu}^{ij}(t)$, and also assume that the time associated with $l^{th}$ repetition of condition $*$ in the fMRI experiment is denoted by $t_l^*$, with total repetitions of the condition $|*|$. Then the neuronal drives associated with population $i$ in region $j$ to the neurovascular function are calculated as follows:

$$z^{ij}(t) = \sum_{l=1}^{|1|} f_1^{ij}(t - t_l^1) + \cdots + \sum_{l=1}^{|nu|} f_{p_p}^{ij}(t - t_l^{nu}) \qquad (11)$$

The $z^{ij}(t)$ in each region are then combined based on the particular hypothesis about neurovascular coupling. In this paper, the neurovascular drives to the haemodynamic response in region $j$ (each region comprises four populations) were calculated using one of the two general forms:

$$\begin{aligned} z^j(t) &= \sum_{i=1}^{4} \beta_{ij} z^{ij}(t) \\ z^j(t) &= f_{nc} \otimes \left( \sum_{i=1}^{4} \beta_{ij} z^{ij}(t) \right) \end{aligned} \quad (12)$$

The first line in equation 12 states that neuronal activity causes the BOLD response instantaneously whereas the second equality introduces a delay and dispersion through the application of a convolution operator that models intracellular calcium dynamics, as in equation 8. We will refer to these two forms as *Direct* and *Delay*, respectively. Parameters $\beta$ are scalars that can be constrained to be identical or vary across regions.

Finally, the third step is to use these neurovascular signals as input to the haemodynamic model of responses in each region or source (see the first line of Equation 9). The parameters and evidence of the haemodynamic models are estimated from the fMRI data using Variational Laplace in the usual way (Friston et al 2007).

## 3. Experimental methods
### 3.1 Example dataset

To illustrate the methods outlined above, we acquired a dataset from a single neurotypical subject (right-handed, male, age 30) who performed the same auditory task while undergoing fMRI and MEG on separate days. This experiment was conducted in accordance with the Ethics Committee of University College London, UCL Ethics Ref: 1825/003 (MRI) and Ref: 1825/005 (MEG).

The task was a variant of the auditory roving oddball paradigm (Baldeweg et al. 2004), which has been extensively characterised in patient and control populations using DCM (e.g. Boly et al., 2012; Dima et al., 2012; Garrido et al., 2008; Rosch et al., 2018). Participants hear a series of 'standard' tones of the same pitch (frequency). Occasionally, the tone changes to a new pitch (a 'deviant'), eliciting neural responses that gradually reduce over the tones that follow, as the deviant becomes the new standard. These neural effects cause marked deviations in the MEG signal (the mismatch negativity, MMN) and we expected there to be concomitant changes in the fMRI signal. We extended the roving oddball task with a second experimental factor of agency. In each block of tones, the auditory stimuli were either produced by the subject ('control' condition) or by the computer ('respond' condition) as detailed in Figure 2.

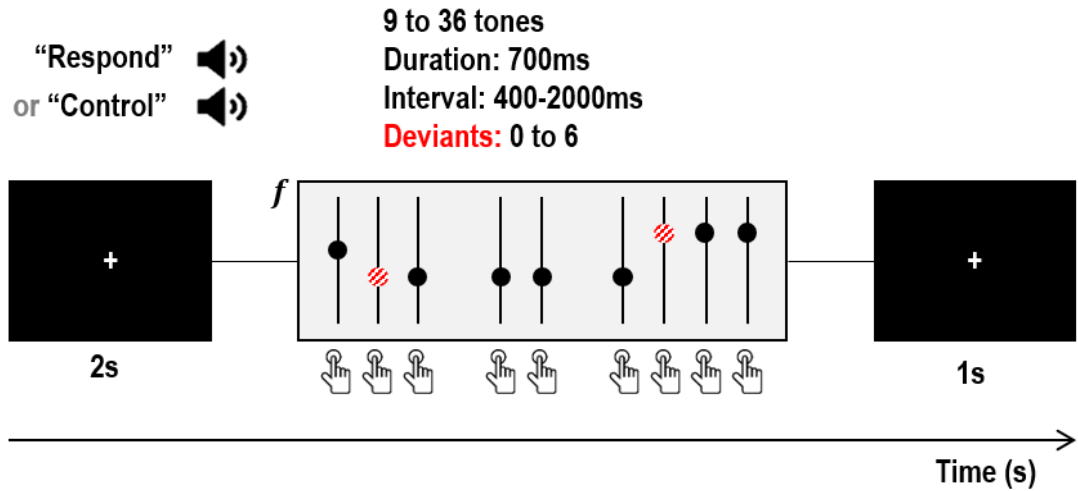

Figure 2. Structure of a single block of the experiment. The subject received an auditory cue, instructing the subject to respond to auditory tones or control the tones (by pressing a button). After 2s, a series of tones was presented. Deviant tones (red striped circles) differed in frequency from the preceding tone. Whether a tone was a standard or deviant was independent of whether the tone was triggered by the computer or the subject. The block ended with an inter-block interval of 1s. Image credits: Press button by Hea Poh Lin and Speaker by ProSymbols from the Noun Project, CC BY 3.0.

There were therefore two independent experimental factors – surprise (standards vs deviants) and agency (computer- vs human-controlled tones). To maximise fMRI efficiency, the auditory stimuli were arranged into blocks of four types – 1) respond with many deviants 2) respond with few deviants 3) control with many deviants 4) control with few deviants. The computer screen in the MRI scanner and MEG system displayed a white fixation cross on a black computer screen, and the subject was instructed to fixate throughout. We will present analyses focussing on the novel manipulation of agency in a separate manuscript. Here, we used data collected under this task purely to illustrate the estimation of neuronal and neurovascular responses in the auditory hierarchy. The MEG and fMRI datasets were pre-processed using standard procedures in SPM12 (for details, see the supplementary material).

### 3.1 Preliminary fMRI analysis

We used the fMRI data to select regions of interest for the subsequent analyses. We specified a General Linear Model with regressors (covariates) encoding the onsets of deviants in the control blocks, deviants in the respond blocks, auditory cues instructing the participant of whether they were in a respond or control block, as well as regressors encoding head motion and a constant term. We computed the t-contrast for the main effect of deviants vs standards, thresholded at $p < 0.05$ family-wise error corrected for multiple comparisons. This identified five regions conventionally included in mismatch negativity studies (Garrido et al.,2008): left and right Heschl's gyri, left and right planum temporale and right inferior frontal gyrus (IFG). We identified the MNI coordinate of the peak response in each region and extracted a single representative timeseries (the first principal component) from each.

## 3.2 DCM for MEG specification

Pre-processing the MEG data gave rise to four types of event-related potential (ERP), namely standards in respond blocks (SR), deviants in respond blocks (DR), standards in control blocks (SC) and deviants in control blocks (DC). We defined a neuronal (CMC) model comprising a fully connected network (by defining priors on adjacency matrix A) to govern dynamics of the four ERP conditions SR, DR, SC and DC in the time interval $[0 - 400]\,ms$ post-stimuli. Differences between the four ERPs were characterised by the following between trial effect (BTF) matrix:

$$BTF = \begin{bmatrix} SR & DR & SC & DC \\ 0 & 0 & 0 & 1 \\ 0 & 0 & 1 & 0 \\ 0 & 1 & 0 & 0 \end{bmatrix}. \tag{13}$$

The $BTF$ matrix instructed DCM for MEG to treat the SR condition as the baseline, and to model each of the remaining conditions by adding condition-specific forward and backward $B$ matrices (Litvak et al., 2011). The priors for the B matrices in this paper were defined such that all extrinsic forward, backward and self-inhibition of neuronal populations were subject to change by the DR, SC and DC conditions. The thalamic inputs, $U$, which encode external stimuli were received by the lowest level in the cortical hierarchy of our model (left and right Heschl's Gyrus), specified by a bell-shaped (Gaussian) function with prior latency of $70 \pm 16\,ms$. We fitted this model to the MEG data using the eight principal modes of the modelled and observed ERPs as data features (Auksztulewicz and Friston 2015; Friston, 2007). Using the posterior expectations of the neuronal parameters, we then used the canonical microcircuit model to simulate neuronal drives (i.e., posterior predictive expectations) for each of the four experimental conditions.

## 3.3 Neurovascular model specification and comparison

The neuronal inputs to the haemodynamic model were generated from the neuronal drive functions, parameterised according to the hypothesis being tested. Let the simulated neuronal response of population $i$ in region $j$ for the four conditions be denoted by $f^{ij}_{SR,DR,DC,SC}(t)$. Using equation 11, the neuronal drives associated with population $i$ in region $j$ to the neurovascular function are given as follows:

$$Z^{ij}(t) = \sum_{l=1}^{|DR|} f^{ij}_{DR}(t - t^{DR}_l) + \sum_{l=1}^{|SR|} f^{ij}_{SR}(t - t^{SR}_l) + \sum_{l=1}^{|DC|} f^{ij}_{DC}(t - t^{DC}_l) \\ + \sum_{l=1}^{|SC|} f^{ij}_{SC}(t - t^{SC}_l) \tag{14}$$

We defined a set of 16 candidate haemodynamic models that covered a number of biologically informed hypotheses about the nature of neurovascular coupling. These models varied according to four model attributes or factors:

Q1: How should neurovascular coupling be parameterised? We considered three options, regarding whether the haemodynamic part of the model should be driven by:

- collaterals from presynaptic inputs to each population, with separate parameters for each population
- collaterals from presynaptic inputs to each population, grouped into excitatory, inhibitory and extrinsic collaterals (Friston et al. 2017)
- postsynaptic neuronal drive ($f$ functions in equation 11)

Q2: Should distal neuronal sources exert changes on the regional BOLD response? I.e. should haemodynamics be driven by local neuronal populations only, or additionally by exogenous inputs from other regions?

Q3: Should neurovascular coupling parameters be region-specific or equal for all regions ($\beta$ in equation 12)?

Q4: Should a Direct or Delay model governing the dynamics of astrocyte responses be used (selection of the first or second equality in equation 12)? This addresses the delays associated with the release of vasoactive agents (e.g., intracellular calcium).

These four questions underwrite 16 candidate models, listed in Table 5. We then estimated the parameters and evidence (free energy) for each of the models using the standard variational Laplace scheme (Friston et al 2007). To address each experimental question, we grouped the candidate models into families and compared them using family-wise Bayesian model comparison (Penny et al., 2010). Finally, we used Bayesian model comparison over the entire model space to find the most parsimonious explanation for the origin of the BOLD response in our dataset.

## 4. Results

We first used the fMRI data to locate brain regions responding to the main effect of deviants versus standards. As hypothesised, this included five regions typically found in the oddball paradigm, shown in Figure 3a.

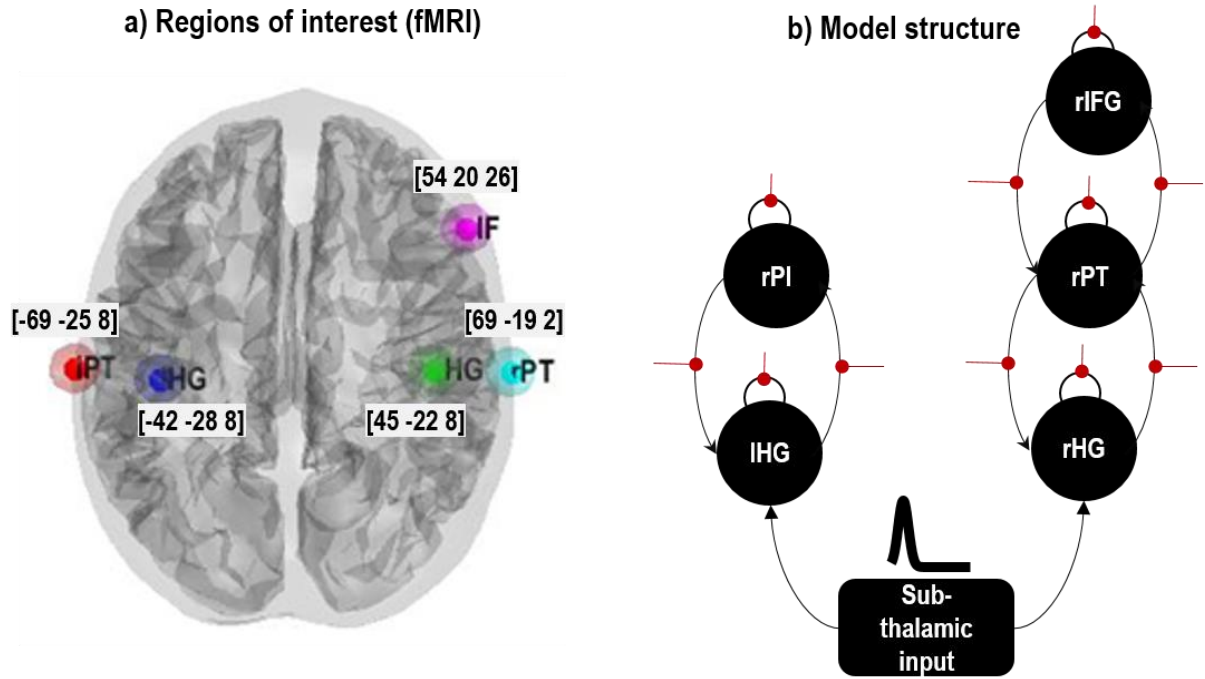

Figure 3. Region of interest selection and DCM network structure. a) Five neuronal sources that were activated during the fMRI experiment, identified using a mass univariate analysis. These were left and right Heschl's gyrus (lHG, rHG), left and right planum temporale (lPT, rPT) and inferior temporal gyrus (rIFG). Peak MNI coordinates, used as priors for MEG source localisation, are shown. b) Structure of the DCM neuronal model. Each large black circle is a canonical microcircuit (CMC), extrinsic connections between regions are shown as curved black lines, and connections that were subject to change from one condition to another are indicated with straight red lines.

Next, we used the coordinates of these five regions as priors for source localisation in DCM for MEG. We specified a DCM, as shown in Figure 3b, where each brain region or source (black circle) was a canonical microcircuit. We fitted this model to the MEG data. Figure 4 shows the scalp maps associated with the prediction of the model and the observed data over the time course of a trial. A close correspondence between the predicted and real data is apparent.

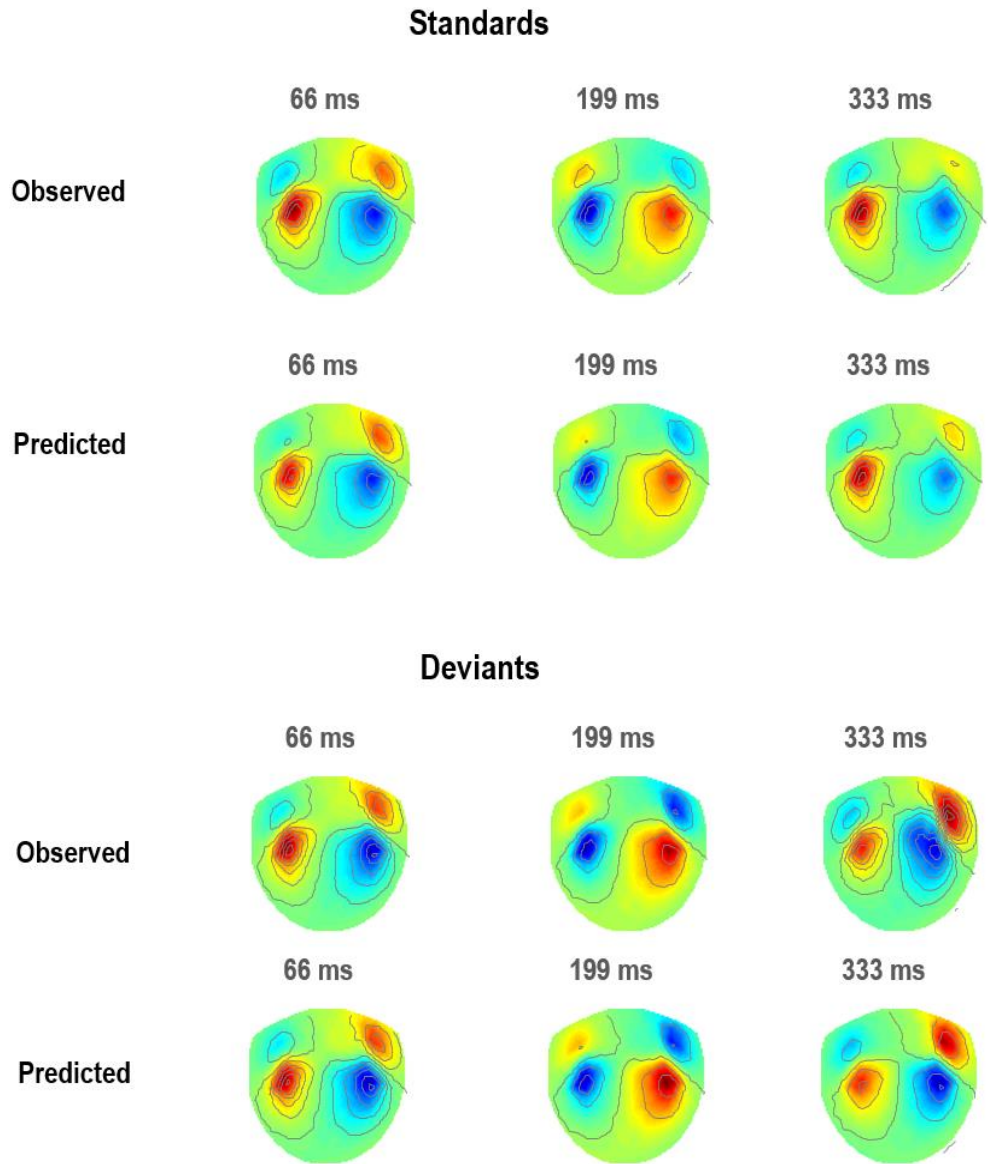

Figure 4. DCM for MEG results. This figure shows scalp map projections of observed and predicted responses for two conditions; namely, standard and deviant tones (in the respond blocks only).

We then used the posterior neuronal estimates to simulate pre/postsynaptic potentials associated with the four experimental conditions – i.e. to generate neuronal drive functions. These are shown in Figure 5a for the inhibitory population in the IFG region (the rest of the neuronal drives were calculated in a similar way). These condition-specific responses were then aligned with the associated conditional onsets in the fMRI experimental design (equation 11 and Figure 5b). Neuronal drives associated with each source were then summed (and in some models filtered to replicate delay dynamics of neurovascular coupling) to generate the neurovascular drive to the haemodynamic model (equation 14 and Figure 5c).

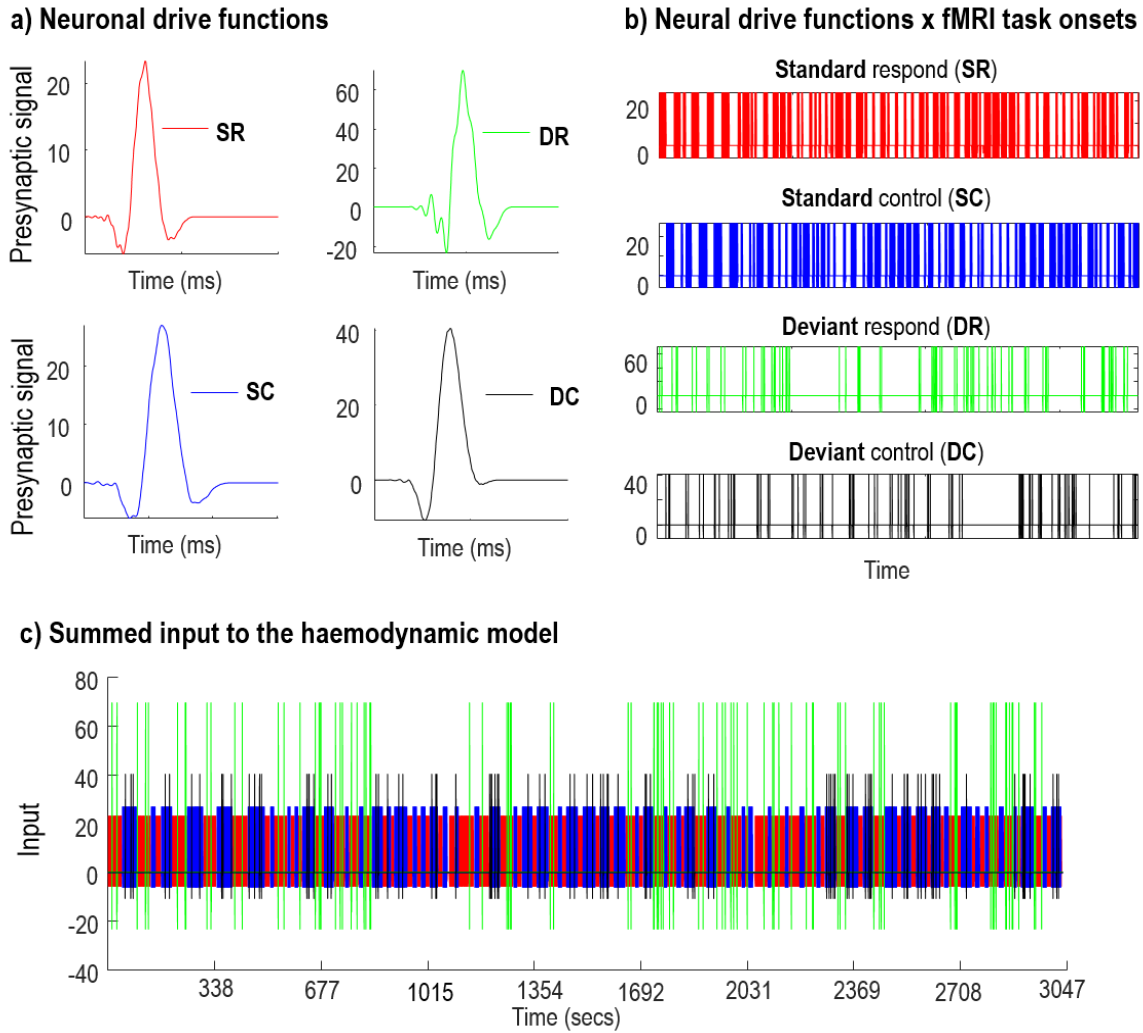

Figure 5. Simulated neuronal drive associated with one neuronal population. DCM for MEG was first used to infer the neuronal parameters of CMC models. a) The ensuing neuronal parameters were used to generate condition specific neuronal responses (e.g., pre synaptic signals). b) To generate the input for the haemodynamic model, the neuronal drive functions were convolved (or shifted in time) with the onset of each trial of the fMRI experiment. c) All condition specific neural responses were then summed to generate the neuronal drives to neurovascular coupling units. This was repeated for each neuronal population and brain region.

As detailed in Section 3.3, we specified and estimated 16 candidate haemodynamic models, which varied in their mechanisms of neurovascular coupling according to four model factors. We then divided the models into 'families' according to each factor and performed a series of family comparisons. The results of Bayesian model comparison showed that neurovascular coupling was best explained (with a posterior confidence approaching 100% for each comparison) as:

(i) driven by collaterals from presynaptic input, separately parameterised for each neuronal population, rather than presynaptic input grouped into excitatory/inhibitory/exogenous connections or postsynaptic input

(ii) driven by local neuronal projections *without* afferent input from distal regions

(iii) separately parametrised on a region-specific basis, rather than having shared weights for each condition and neuronal populations across brain regions

(iv) having a direct form of model governing the dynamics of astrocyte responses, as opposed to a delayed effect.

The overall winning model, with a log Bayes factor of 7.67 compared to the next best model, suggested the BOLD response is driven by instantaneous local presynaptic neuronal activity, with region-specific parameterisation of neurovascular coupling. Figure 6 shows the estimated neurovascular coupling parameters from this model, with parameters not contributing to the model evidence pruned using Bayesian model reduction. For each parameter, Bayesian model reduction was used to test the hypothesis that the parameter was present vs absent (i.e. non-zero vs zero). In this plot, each group of four bars are the estimated contribution of each neuronal population (SS, SP, II, DP) to the haemodynamic model. We will not attempt to draw strong conclusions from this result, as only data from an exemplar subject was used. Nevertheless, in all five regions there were parameters which deviated confidently from their prior expectation of zero, confirming that the synaptic activity estimated from the MEG data captures variance in the fMRI data (explained variance per region: 53%, 37%, 64%, 37% and 28%). The next step will be to determine whether the results of the illustrative model comparisons presented above, as well as the estimated parameter values (e.g. the pattern of positive and negative parameters) replicate across subjects in larger group studies and under different experimental paradigms.

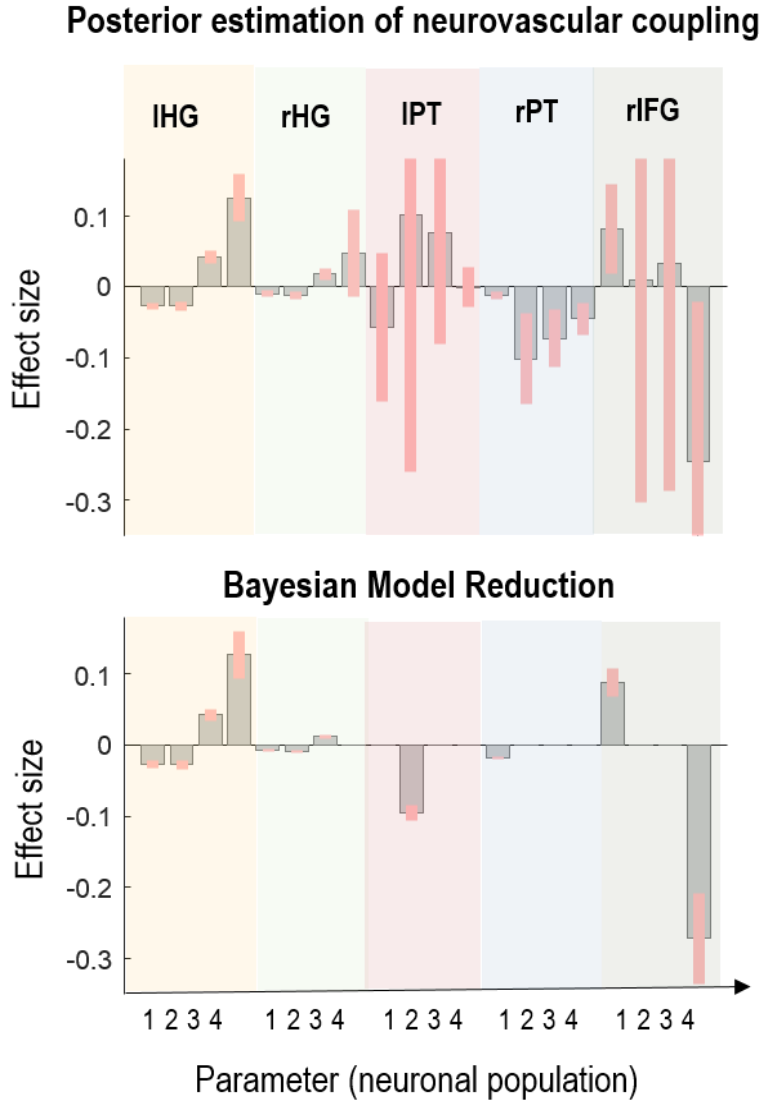

Figure 6. Estimated neurovascular parameters. Posterior estimates of the neurovascular coupling parameters β that best accounted for the multimodal data and BMR analysis of estimated parameters that elucidate key parameters governing dynamics of data. The grey bars are the expected values and the pink error bars are 90% credible intervals. Each group of four bars corresponds to parameters quantifying the contribution to the neurovascular coupling by: spiny stellate (SS), superior pyramidal (SP), inhibitory interneurons (II) and deep pyramidal (DP) cells. The titles indicate the brain regions: left Heschl's gyrus (lHG), right Heschl's gyrus (rHG), left planum temporale (lPT), right planum temporale (rPT), right inferior frontal gyrus (rIFG).

## 5. Discussion

The novel contribution of this work is to establish a relatively straightforward multi-modal DCM approach that flexibly connects laminar-specific neural mass models, which are fitted to electrophysiological data, with neurovascular models, which are fitted to fMRI data, via simulated neuronal drive functions. Together, these form a complete generative model of the BOLD signal, which enables hypotheses about neurovascular coupling to be tested efficiently using Bayesian model comparison and reduction. The neuronal drive functions act as a bridge between the fMRI and MEG modalities, enabling multi-modal analyses to be conducted with any of the neural mass models implemented within the DCM framework. We addressed the difficult parameter identification problem inherent in having a single generative model of both BOLD and MEG signals (e.g. Friston et al., 2017) by separately estimating neuronal parameters using MEG data, and neurovascular / haemodynamic parameters using fMRI data. This can be seen as a simple form of Bayesian belief updating, in which the posterior estimates based upon MEG data are used as a precise priors for models of haemodynamic responses, which share a common set of neuronal parameters. Notice that using MEG data to inform the characterisation of fMRI data rests explicitly on having a common DCM that can generate both modalities – and that share the same neuronal parameters and architecture. Crucially, we can leverage this form of Bayesian belief updating using 'off the shelf' dynamic causal models for both modalities. The only thing we need to add is a neuronal drive function that links the modality-specific DCMs. The proposed approach may offer new insights into the source of the BOLD signal in the healthy and pathological brain and is available through the SPM software.

To illustrate the type of questions that can be addressed using this approach, we used Bayesian model comparison to address four questions in a single subject MEG/fMRI dataset. It should be emphasised, that given this was only a single subject dataset, these examples should be regarded as a proof of concept rather than a definitive result.

The first question was whether BOLD signal was best explained as being driven by a Direct (scaling only) or Delay model (scaling and delay) of neurovascular coupling. This question was motivated by studies in animal models, suggesting a delay between neuronal activity and the BOLD response caused by elevation of intracellular calcium in astrocytes collaterals (Rosenegger et al., 2015). We used a lumped linear second order model, which can be effectively inferred using fMRI data. Such a model component is simple and effective (parsimonious), given the limited temporal resolution of fMRI data. Our analysis suggested instantaneous electrophysiological fluctuations induce BOLD responses directly, which agrees with Logothetis (2003).

The second question we addressed was whether presynaptic or postsynaptic neuronal activity mediated haemodynamic responses. Based on this single subject, our analysis showed that BOLD is likely to be caused by presynaptic signals. This is in line with the findings of Attwell and Iadecola (2002) and Logothetis (2003, 2008), who concluded that mean neuronal firing rates (presynaptic signals) are largely responsible for the BOLD response.

The third question we addressed was whether extrinsic collateral afferents from distal regions contribute to haemodynamics, or whether neurovascular coupling should be considered a purely local phenomenon. Bayesian model comparison suggested that local neuronal activity provided the best explanation for BOLD response, as is assumed, for example, in mass-univariate (SPM) analysis or Dynamic Causal Modelling (DCM) for fMRI.

The final question we addressed was whether contribution of neuronal populations to the neurovascular units were identical across brain regions or region specific. We found strong evidence for the former – a single parameterisation across brain regions provided the best explanation for the data. This agrees with physiological wisdom that neurovascular mechanisms are not only region specific in the brain (Devonshire et al. 2012), but also different across cortical layers (Goense et al. 2012 & 2015).

This framework may be particularly useful for studying processes that effect both neuronal and haemodynamic responses. For instance, it could be used to model effects of aging (D'Esposito et al., 2003) in cognitive paradigms, where the factor of aging would be expected to not only affect neuronal responses, but also stiffness of blood vessels, quantified by Grubb's exponent in the Balloon model (see Equation 8) and/or delays in the model neurovascular coupling. To facilitate this, multimodal DCM could be combined with the parametric empirical Bayes method (Friston et al., 2016), to test for differences in neurovascular and haemodynamic parameters between young and old age groups. The proposed approach in this paper may also be useful for characterising experimental manipulations for which neurovascular function alone is altered. For instance, the action of a particular intervention such as diazoxide is predominantly on neurovascular coupling, with little effect on neuronal dynamics (Pasley 2008). Potentially, our proposed approach, together with PEB for random effects analysis, are well placed to characterise and elucidate neurovascular physiology.

The approach described here affords the opportunity to investigate laminar-specific contributions to the BOLD signal, without requiring laminar-specific fMRI data. However, a key limitation of the model is the assumption of a single haemodynamic compartment. In fact, neural vasculature has a well-studied spatial arrangement in the cortical depth, which was modelled in the DCM framework by Heinzle et al. (2016). This could be incorporated in the approach described here, in order to better account for differences across laminae due to vasculature. Furthermore, as high spatial resolution fMRI data becomes more readily available – with the rollout of 7-tesla scanning – the question arises of how to make use of these data for informing estimates of neurovascular coupling parameters. There is considerable interest in associating the BOLD response to specific layers of the cortical column, and associated with this, top-down and bottom-up connections (e.g. Scheeringa & Fries., 2017, Lawrence et al., 2017, Duyn, 2012). Typically, laminar fMRI involves dividing the cortical depth into two or three layers and extracting timeseries from each. Incorporating these into the framework presented here could, in principle, be achieved by incorporating a mapping between layers in the data and cortical layers in the haemodynamic model.

In summary, we hope the statistical tools presented here will prove useful, both for the ongoing development of neurovascular models and the application of these models for testing hypotheses using multi-modal data. We provide code and example data via SPM.

## 6. Acknowledgments

The Wellcome Centre for Human Neuroimaging is supported by core funding from Wellcome [203147/Z/16/Z]. We are grateful to Marta Garrido for helpful conversations about the design of the oddball task and to Alphonso Reid and Clive Negus for their support with data collection.

## 8. Tables

**Table 1:** Parameters of the neuronal model (see also Figure 2).

| | Description | Parameterisation | Prior |
|---|---|---|---|
| $\kappa_i$ | Postsynaptic rate constant of the $i$-th neuronal population in each of $N$ regions | $\exp(\theta_\kappa) \cdot \kappa_i$ <br> $\kappa = [256, 128, 16, 32]$ | $p(\theta_\kappa) = N(0,0)$ |
| $a_{i \to k}$ | Intrinsic connectivity between populations $i$ and $k$ in each region. | $\exp(\theta_a) \cdot a$ <br> $a = [2\ 1\ 1\ 1] * 512$ | $p(\theta_a) = N(0,0)$ |
| $B_{b,f}$ | Conditional-specific matrix. Priory entries of these matrix are zero unless forward, backward or intrinsic connections are allowed to change in different conditions. | $\theta_{b,f}$ | $p(\theta_b) = N(0, \frac{1}{8})$ |
| $A_{f,b}$ | Forward and backward extrinsic connectivity matrices. If there is any forward (backward) connection between from region $j$ to $i$, the corresponding element $(i,j)$ in $A_f$ ($A_b$) is set to one. | $\exp(\theta_A) \cdot A_{f,b}$ | $p(\theta_A) = N(0, \frac{1}{8})$ |
| $C$ | Scaler matrix to driving input | $\theta_c$ | $p(\theta_C) = N(0, \frac{1}{32})$ |

**Table 2**: Parameters of neurovascular and haemodynamic responses function.

| | Description | Parameterisation | Prior |
|---|---|---|---|
| $\eta$ | Rate of signal decay per sec | $0.64 \cdot \exp(\theta_\eta)$ | $p(\theta_\eta) = N(0, \frac{1}{256})$ |
| $\chi$ | Rate of flow-dependent elimination | $0.32 \cdot \exp(\theta_\chi)$ | $p(\theta_\chi) = N(0,0)$ |
| $\tau$ | Rate hemodynamic transit per sec | $2.00 \cdot \exp(\theta_\tau)$ | $p(\theta_\tau) = N(0, \frac{1}{256})$ |
| $\alpha$ | Grubb's exponent | $0.32 \cdot \exp(\theta_\alpha)$ | $p(\theta_\alpha) = N(0,0)$ |
| $\varepsilon$ | Intravascular : extravascular ratio | $1.00 \cdot \exp(\theta_\varepsilon)$ | $p(\theta_\varepsilon) = N(0, \frac{1}{256})$ |
| $\varphi$ | Resting oxygen extraction fraction | $0.40 \cdot \exp(\theta_\varphi)$ | $p(\theta_\varphi) = N(0, 0)$ |
| $\beta_i$ | Sensitivity of signal to neural activity | $\theta_\beta$ | $p(\theta_i) = N(0, \frac{1}{16})$ |
| $\tau_{nc}$ | Decay rate of the astrocytes collateral | $0.7 \cdot \exp(\theta_{\tau_{nc}})$ | $p(\theta_{nc}) = N(0, \frac{1}{16})$ |

**Table 3**: Biophysical parameters.

| | Description | Value |
|---|---|---|
| $V_0$ | Blood volume fraction | 0.08 |
| $k_1$ | Intravascular coefficient | $6.9 \cdot \varphi$ |
| $k_2$ | Concentration coefficient | $\varepsilon \cdot \varphi$ |
| $k_3$ | Extravascular coefficient | $1 - \varepsilon$ |

**Table 4:** Glossary of variables and expressions.

| Variable | Description |
|---|---|
| U | Sub thalamic Gaussian shape function. |
| $V_j^K$ | The $j$-th (neuronal) state in region $k$; e.g., mean depolarisation of a neuronal population |
| $\sigma(V_j^k)$ | The neuronal firing rate – a sigmoid squashing function of depolarisation |
| z | Neurovascular signal; e.g., intracellular astrocyte calcium levels |
| $h_s, h_{in}, h_v, h_q$ | Haemodynamic states: $h_s$ - vasodilatory signal (e.g., NO), $h_{in}$ - blood flow, $h_v$ - blood $h_q$ - volume deoxyhaemoglobin content |
| $\Psi_j$ | Electromagnetic field vector mapping from (neuronal) states to measured (electrophysiological) responses |

**Table 5:** Model space design to investigate function of neurovascular coupling.

| Model | F1: Parameterization | F2: Distal inputs? | F3: Region-specific? | F4: Direct vs Delay |
|---|---|---|---|---|
| 1 | Pre | Yes | Yes | Direct |
| 2 | Pre | No | Yes | Direct |
| 3 | Pre | Yes | No | Direct |
| 4 | Pre | No | No | Direct |
| 5 | Post | N/A | Yes | Direct |
| 6 | Post | N/A | No | Direct |
| 7 | Pre (Friston et al.2017) | Yes | No | Direct |
| 8 | Pre (Friston et al.2017) | No | No | Direct |
| 9 | Pre | Yes | Yes | Delay |
| 10 | Pre | No | Yes | Delay |
| 11 | Pre | Yes | No | Delay |
| 12 | Pre | No | No | Delay |
| 13 | Post | N/A | Yes | Delay |
| 14 | Post | N/A | No | Delay |
| 15 | Pre (Friston et al.2017) | Yes | No | Delay |
| 16 | Pre (Friston et al.2017) | No | No | Delay |

* Factors F1-F4 correspond to the factors of the experimental design described in Section 2.3

# Supplementary material

*Neurovascular coupling: insights from multi-modal dynamic causal modelling for fMRI and MEG*

*Jafarian et al.*

## Supplementary methods

### Participants

We scanned a single subject (male, right-handed, age 30) performing the same task under fMRI and MEG. This experiment was conducted in accordance with the Ethics Committee of University College London, UCL Ethics Ref: 1825/003 (MRI) and Ref: 1825/005 (MEG).

### Task

This study used a novel version of the auditory roving (mismatch negativity) oddball paradigm (Garrido at al. 2008) which included an additional factor of agency, such that the auditory stimuli were produced by the subject (self) or by the computer (other). The subject alternated between responding with button presses to a series of computer-generated tones, and generating a series of tones himself using button presses. During the alternation of subject- and computer-generated sequences, the tone of the stimulus changed sporadically, producing oddball responses – that resolved over trials as the new tone became the standard.

There were two experimental factors – surprise (standards vs deviants) and agency (computer- vs human-controlled tones). To maximise fMRI efficiency, the subject was presented with auditory stimuli arranged into blocks, and there were four block types – 1) respond with many deviants 2) respond with few deviants 3) control with many deviants 4) control with few deviants. The experiment was entirely auditory – the computer screen in the MRI scanner displayed a white fixation cross on a black computer screen, and the subject was instructed to fixate throughout.

The structure of a block is summarised in Figure 2 of the main text of the paper. At the start of computer-controlled blocks, the subject heard the auditory cue 'respond'. A sequence of 70ms auditory tones was then presented with irregular intervals between them, ranging from 400ms to 2000ms. The subject pressed their button each time they heard a tone. The human-controlled blocks started with the auditory cue 'control'. The subject controlled the onset of the tones, and were instructed to press their button to trigger tones at times of their choosing. They were trained to keep the pace of button-pressing similar to the computer-controlled blocks, but to freely alter the time between individual tones as desired. To ensure that there were no systematic differences in the intervals across conditions, the intervals between tones in the computer-controlled block $n$ were taken from human-controlled block $n - 2$ and their order was reversed, to reduce the possibility of the timing sequence being recognisable.

The tones within a block had auditory frequencies 500Hz, 600Hz, 700Hz and / or 800Hz. A 'deviant' tone was one which differed in frequency from the previous tone, whereas a 'standard' tone had the same frequency as the previous tone. The first tone in a block always had the same frequency as the last tone of the previous block. For the 'many deviant' blocks, the number of deviants was sampled from a Poisson probability density function $f$ and calculated using the following equation:

$$n = 1 + f(x|\lambda = 1.5)$$

Where $x$ was a random vector. In the 'few deviant' blocks, all but two blocks per run had zero deviants (i.e. all standards), and two blocks had one deviant.

The number of tones in each block was varied, to reduce anticipation of the end of the block. In the many deviant blocks there were 28, 30, 32, 34 or 36 tones, and in the few deviant blocks there were 9, 10, 11, 12 or 13 tones. By reducing the number of tones in the few deviant blocks, the design efficiency was improved (by avoiding unnecessary over-sampling of the standard tones).

The subject was instructed on how to perform the task (see Supplementary text: subject Briefing) and performed a practice run in front of a desktop PC, consisting of 4 blocks. They were then positioned in the MRI scanner and performed a further practice run of 4 blocks. They then performed 3 runs of the task while undergoing fMRI, where each run consisted of 860 tones divided across 40 blocks. There were 56, 54 and 64 deviant tones in each run respectively. The subject then returned on a separate day and repeated the first two runs of the experiment in the MEG (which differed only in the human-controlled stimulus timings).

### Stimulus presentation and onset timing identification.

The experiment was controlled using the Cogent2000 software (http://www.vislab.ucl.ac.uk/cogent_2000.php). In both the MRI and MEG scanners, auditory stimuli were triggered using a low-latency audio presentation system (AudioFile Stimulus Processor, Cambridge Research Systems, Rochester, UK) and delivered to the subject using the Ear-Tone Etymotic stereo sound system (Etymotic Research Inc., Illinois, USA). The timing of auditory stimuli and button presses were recorded in the MRI scanner using a Micro 1401 Mk II connected to a computer running the Spike2 software version 6 (Cambridge Electronics Devices, Cambridge, England).

There were four experimental conditions in the paradigm, and the onsets and offset times of each trial needed to be identified precisely for the analyses presented here. These were ($i$) standard tones in the control block, ($ii$) deviant tones in the control block, ($iii$) deviant tones in the respond block and ($iv$) standard tones in the respond block. For simplicity, the first tone of a new frequency or block was defined as a deviant, and all other tones were defined as standards. The onset of each trial was detected from the Spike2 audio recording, in order to account for any latency in audio presentation. This detection was based on a pre-defined threshold, and the identity of the experimental condition was determined from the auditory cues using a classification method that utilized the dynamic time warping (DTW) algorithm. To clarify this, first we saved two separate audio files for the words 'control' and 'respond' from the audio files (hereinafter called templates). Next, using the pre-defined threshold, we detected the onset and offset of each word in the timeline of the experiment. These onsets/offsets were employed to detect the timing of each individual word. Each individual word was then compared with the two audio template using dynamic time warping (DTW) distance. A word in the experiment timeline was identified as control (respond) if its DTW distance to the template audio file of the word 'control' ('respond') was smaller than 'respond' ('control').

### MRI data acquisition & preprocessing

A 2D gradient echo echo-planar-imaging sequence was used to acquire the functional data in a 3T Prisma scanner (Siemens Healthineers, Erlangen, Germany). A 64 channel coil was used for signal reception and an integrated body coil for transmission. Each volume consisted of 48 transverse slices and was acquired in 3.36s, with ascending slice order. The following parameters were used: voxel size of 3 mm x 3 mm in-plane; slice thickness of 2.5mm, 0.5mm slice separation, field of view of 192 mm × 192 mm, 12% over-sampling in the phase-encoded direction, bandwidth of 2298 Hz/px, echo spacing 0.5ms, echo time of 30 ms; flip angle of 90°. Fat saturation with an excitation of 130deg was used prior to each excitation. A fieldmap was also acquired to correct the images for distortions.

A T1-weighted MPRAGE (magnetisation-prepared 3D rapid gradient echo) anatomical image (Mugler and Brookeman MRM '90) was acquired with the following parameters: inversion time was set to 1100 ms; excitation flip angle was 7°; time to echo was 3.34 ms; receiver bandwidth was 200 Hz/pixel; echo spacing was 7.4 ms; and repetition time, 2530 ms. Reconstruction matrix dimensions were 256×256×176, with 1×1×1 mm3 voxel size. Parallel imaging acceleration (Griswold et al. MRM '02) was enabled with acceleration factor of 2 and 32 integrated reference lines, giving a scan time of 6:03 minutes.

Functional and structural images were bias-corrected and then pre-processed using the standard pipeline in SPM12 (revision 7265). Functional images were realigned and unwarped using the acquired fieldmaps and the structural image was segmented into constituent tissue types. The functional images were co-registered to the structural, and all images were normalised to MNI space. Explicit smoothing with a 6mm FWHM kernel was applied to the functional images, in order to increase SNR and to satisfy the requirements of multiple comparisons correction using random field theory.

The preprocessed fMRI data were analysed by specifying a single General Linear Model (GLM) for each subject, concatenated over runs. This included regressors for the deviants in the respond condition, deviants in the control condition, cues, physiological regressors derived from breathing / pulse measurements and a constant term for each run. The GLM was estimated and results assessed at a voxel-wise threshold of p < 0.05 family-wise error (FWE) corrected, shown in Figure S1 and Table S1.

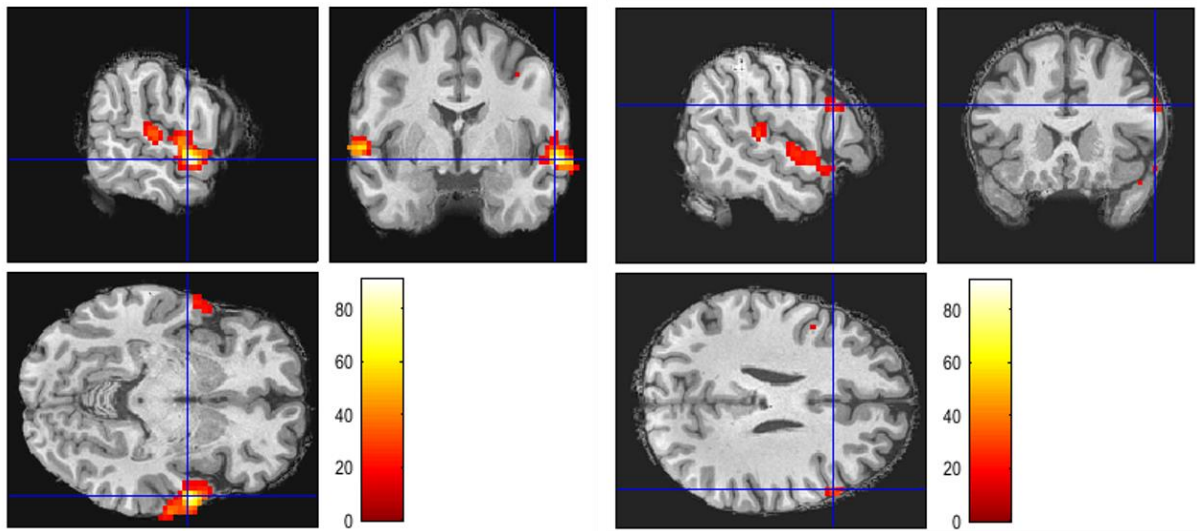

Figure S1. Functional MRI results. The panels show the thresholded Statistical Parametric Mapping (SPM) results with sagittal and coronal planes on the top row and the axial plane on the bottom row. Left: the crosshair positioned on right planum temporale (rPT), MNI coordinates: [69, -19, 2]. Right: the crosshair positioned on right inferior frontal gyrus (rIFG), MNI coordinates: [54, 20, 26].

Regions of interest were identified by positioning spheres of radius 8mm at the coordinate of peak activation of each region (listed in Table S1). Each region's timeseries were summarised over significant voxels (p < 0.05 FWE corrected) by taking their first principal component (eigenvariate). Each summary timeseries was high pass filtered, pre-whitened, and corrected for known confounds (the mean and physiological regressors). These timeseries were used for the haemodynamic modelling, below. An additional GLM was specified including all standards and deviant tones, which provided the timing information needed for the haemodynamic modelling.

MEG data acquisition & pre-processing

MEG recordings were made using a 275-channel Canadian ThinFilms (CTF) MEG system with superconducting quantum interference device (SQUID)-based axial gradiometers (VSM MedTech, Vancouver, Canada) in a magnetically shielded room. The data collected (included button presses and onsets of audio stimuli) were digitized continuously at a sampling rate of 600 Hz.

The data were first epoched into different segments. Each segment was the timeline of one of the experimental conditions; namely a standard tone in a control block, a deviant tone in a control block, a deviant tone in a respond block and a standard tone in a respond block. Each segment was high and low pass filtered in the range of [0.5 35] HZ, respectively. Then we use the conventional averaging method to calculate evoked responses associated with each stimulation within a range of [0 400] ms for each individual condition. The ensuing evoked responses for different conditions and channels are shown in Figure S2.

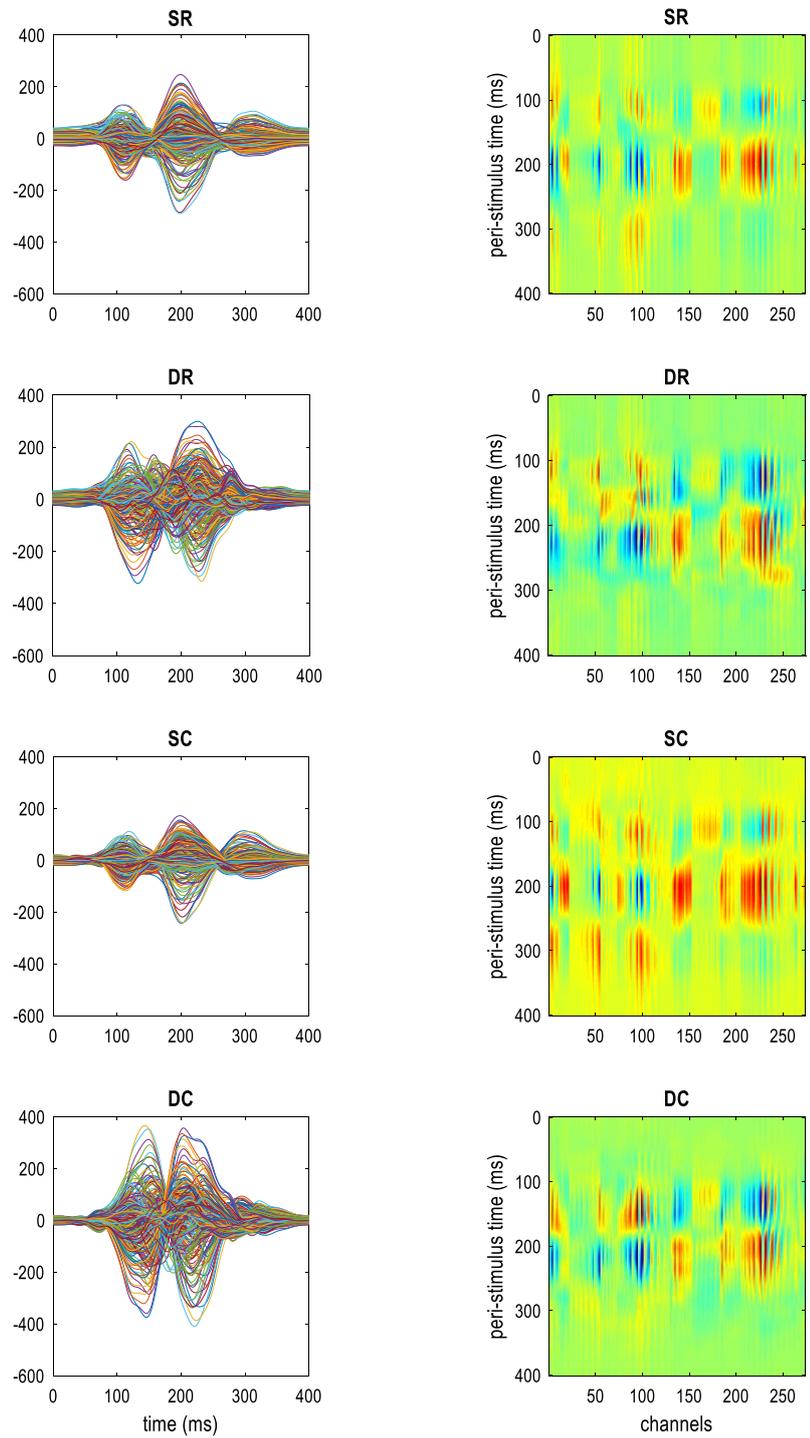

Figure S2 Average evoked responses from the MEG data. The left hand side shows traces of evoked responses over all channels associated with standard tones in respond blocks (SR), deviant tones in respond blocks (DR), standard tones in control blocks (SC), and deviant tones in control blocks (DC). The right hand side shows the heat map of changes of brain activity over different channels for SR, DR, SC and DC evoked responses.

## Supplementary table 1: SPM results

| Contrast | Region | T-statistic (peak) | Coordinates (xyz) |
|---|---|---|---|
| Deviants - Standards | Right planum temporale | 13.48 | 69 -19 2 |
| Deviants - Standards | Left planum temporale | 9.49 | -69 -25 8 |
| Deviants - Standards | Right IFG | 8.16 | 54 20 26 |
| Auditory cues | Right HG | 7.47 | 45 -22 8 |
| Auditory cues | Left HG | 8.97 | -42 -28 8 |

\* Results were computed at p < 0.05 FWE-corrected, limited to a priori regions of interest based on previous studies.

## Supplementary text: subject briefing

The wording used to brief the subject before scanning was as follows:

> *This experiment is all about investigating how we hear sound. In the scanner, you'll be wearing headphones and holding some buttons which you can press.*
>
> *Sometimes you'll hear the computer say the word "respond", and then you'll hear some beeps. Your task is simply to press a button every time you hear a beep. Try and press it as soon as you can when you hear a beep.*
>
> *At other times, the computer will say the word "control". Now, you're in control of the beeps. Every time you press a button, you'll hear a beep. I'd like you press the button at a similar speed as the beeps you heard the computer making when you were responding. But I'd like you to mix it up a bit – sometimes press the button a bit faster, sometimes press it a bit slower. You are in control.*
>
> *So it's as simple as that. When the computer says "respond", press your button when you hear a beep. When the computer says "control", press your button at your own pace. Keep looking at the cross in the middle of the screen during the task (feel free to blink). Any questions?*